\documentclass[12pt, aps, prb, preprint]{revtex4}

\usepackage{graphicx}
\usepackage{amssymb}
\usepackage{endnotes}

\begin{document}
\title{Comparing the efficacy of multimedia modules with traditional textbooks for learning introductory physics content}

\author{Timothy Stelzer}

\author{Gary Gladding}

\author{Jos\'e Mestre}

\author{David T. Brookes}

\affiliation{Department of Physics; University of Illinois at Urbana-Champaign; Urbana, IL 61801}

\date{\today}

\begin{abstract}

A clinical study was performed comparing the efficacy of multimedia learning modules with traditional textbooks for the first few topics of a calculus based introductory electricity and magnetism course.  Students were randomly assigned to three different groups experiencing different presentations of the material; one group received the multimedia learning module presentations and the other two received the presentations via written text. All students were then tested on their learning immediately following the presentations as well as two weeks later. The students receiving the multimedia learning modules performed significantly better than the students experiencing the text-based presentations on both tests.
 
\end{abstract}

\maketitle

\section{Introduction}

\subsection{The instructional dilemma in introductory physics}

How to best spend lecture time to promote student learning in introductory courses is a topic that we often debate with ourselves and with our colleagues.  Recent research suggests that actively engaging students in learning activities results in better conceptual understanding\endnote{R. R. Hake, ``Interactive-engagement versus traditional methods: A six-thousand-student survey of mechanics test data for introductory physics courses,'' Am. J. Phys. {\bf 66}, 64--74 (1998).}  compared to passively listening to lectures.  There is also evidence that students are better prepared to learn difficult concepts from lectures\endnote{D. L. Schwartz and J. D. Bransford, ``A Time for Telling,'' Cogn. \& Instr. {\bf 16}, 475--522 (1998).} or lecture demonstrations\endnote{D.R. Sokoloff and R. K. Thornton, ``Using interactive lecture demonstrations to create an active learning environment,''  Phys. Teach. {\bf 35}, 340--347 (1997).; D.R. Sokoloff and R. K. Thornton, ``Learning motion concepts using real-time microcomputer-based laboratory tools,''  Am. J. Phys. {\bf 58}, 858--867 (1990).} when they are aware of what they don't know and don't understand.  The common thread running through this diverse set of studies is that there is more and deeper learning when students come to class with some basic exposure to the content so that they are prepared to learn from the experiences in lecture.  Reading the textbook prior to coming to class is one way of providing the necessary basic exposure to content, but getting students to read the textbook is a difficult goal to achieve.

Data from a survey administered to students at the University of Illinois at Urbana-Champaign (UIUC) over three consecutive years, shown in Table~\ref{textbooksurvey}, indicate that both life-science majors and engineering majors seldom, if ever, open the textbook prior to coming to class.  About 70\% of both groups admit they ``rarely'' or ``never'' read the textbook;  69\% of both groups claim that the textbook was either ``useless'' or ``not very useful'' in helping them understand the course material. What makes these findings surprising is that at UIUC we attempt to enforce textbook reading prior to class time by administering Just-in-Time Teaching (JiTT)\endnote{G. M. Novak, E. T. Patterson, A. D. Gavrin, and W. Christian, {\it Just-In-Time-Teaching} (Prentice Hall, Upper Saddle River, NJ, 1999).}  ``preflight'' questions crafted from the readings.  These data indicate that students answer the preflight questions based on intuition, or merely to fulfill a requirement, rather than from an informed perspective.  Based on the comparability of UIUC students to students nationwide, and on the frequent complaints voiced by physics professors across the country that students come to class unprepared, we believe that students' ``textbook irrelevance'' attitude is quite prevalent.

These circumstances create a dilemma.  On the one hand, research suggests that we should actively engage our students during class and that instructors' expertise is best used to coach students in learning the subtleties of physics concepts and problem solving; but this requires that students come to class with some exposure to basic content and with some idea of what they do and do not understand---something that we know students are not doing.  On the other hand, recognizing that students come to class unprepared to participate in meaningful activities suggests that we should spend class time covering basic content from the textbook; but this would leave little time for coaching students in developing conceptual understanding and problem solving skills.

\subsection{Multimedia learning}

One possible solution to this dilemma is to use multimedia presentations to help students learn basic content prior to coming to class.  There are three reasons why multimedia presentations might prove effective in this context.  First, today's generation of students has grown up in a multimedia age, and they not only know how to use multimedia better than the current generation of physics professors, but they also like using multimedia.  Second, presenting basic physics content via asynchronous, web-based multimedia presentations can be monitored and enforced better than policing textbook reading.  Multimedia modules can be designed with embedded assessments, and a modest amount of course credit can be assigned to viewing the materials (so long as some minimum time is spent on them) and to performance on the embedded assessments.\endnote{Although the same can be done with JiTT, students resent being ``quizzed'' on textbook reading prior to each lecture, so we tend to simply give credit for answering the questions.  Nothing is to prevent the same thing from happening with multimedia materials, but students' prefence to learning from multimedia compared to textbooks, as demonstrated in our survey question presented in Sec.~\ref{3C}, suggest otherwise.}

The third reason why multimedia presentations of basic content might provide a viable solution to the textbook-irrelevance problem is that a considerable body of research exists on how to design multimedia materials to improve learning.  This is in stark contrast to the situation with textbooks, where enhancements continue to be added (e.g., addition of color; boxes containing problem solving tips, reflection questions, and questions to quiz oneself) with no evidence on whether they add, or perhaps detract, from student learning.\endnote{Although there has been work by psychologists on science text comprehension (see for example, J. Otero, J. A. Leon, and A. C. Graesser, {\it The Psychology of Science Text Comprehension} (Lawrence Erlbaum Associates, Mahwah, NJ, 2002).), the issues of focus have been broad and theoretical. Little research exists comparing learning from textbooks to other forms of presentations at the college level, or examining what features of textbooks enhance, or detract from learning.}  Principles for designing multimedia materials draw from both cognitive science and empirical studies.  For example, work on cognitive load over the last twenty years by Sweller and his collaborators\endnote{R. C. Clark, F. Nguyen, and J. Sweller, {\it Efficiency in Learning: Evidence-based Guidelines to Manage Cognitive Load} (Pfeiffer, 2006).}  indicates that if the student is using valuable working memory resources in order to simply coordinate images and words, s/he has less capacity for actual learning.\endnote{J. Sweller, P. Chandler, P. Tierney, and M. Cooper, ``Cognitive load and selective attention as factors in the structuring of technical material,'' J. Exp. Psych.: Gen. {\bf 119}, 176--192 1990.; R. Tarmizi and J. Sweller, ``Guidance during mathematical problem solving, J. Educ. Psych. {\bf 80}, 424--436 1988.; M. Ward and J. Sweller, ``Structuring effective worked examples,'' Cogn. \& Instr. {\bf 7}, 1--39 1990.}  The expanding research base in the field of multimedia learning has led to the following conclusions:\endnote{Much of the original literature has been collected in: {\it The Cambridge Handbook of Multimedia Learning} (Cambridge University Press, 2005), edited by R. E. Mayer.\label{mayer2005}} {\it a)} people have separate audio and visual channels, {\it b)} these channels have limited capacity, and {\it c)} learning involves the active selection, organization and integration of the information presented via the auditory and visual channels.\endnote{R. E. Mayer, {\it Multimedia Learning} (Cambridge University Press, 2001).\label{mayer2001}}   

Several design principles have emerged from multimedia learning research findings.  For example, students learn better {\it (i)} from words and pictures than from words alone, {\it (ii)} when words and pictures are presented simultaneously rather than successively, {\it (iii)} when extraneous words, pictures and sounds are excluded, and {\it (iv)} from animation and narration than from animation and on-screen text. Although most of these are intuitive, one principle that is perhaps somewhat counterintuitive is that students learn better from animation and narration than from animation, narration and on-screen text.  Overall, this body of research suggests that in designing multimedia learning materials the goal should be to stay on-message with respect to the learning goals and to use different input channels (visual, auditory) to provide complementary but necessary information to help the learner build meaning and understanding.

\subsection{Question to be addressed}

The current situation can be summarized as follows:  Students don't read the textbook as much as we would like and textbook reading has proven difficult to enforce.  On the other hand, viewing and answering embedded questions in multimedia presentations delivered via the web could be more closely monitored and awarded credit, encouraging students to conduct the requisite pre-lecture preparation.  There is, however, no research at the university level examining whether or not students learn basic physics content better from multi-media presentations compared to reading the textbook.  Before any major effort is devoted to designing and implementing multimedia presentations in large enrollment courses, we need to know if they are at least as effective as reading a textbook in terms of imparting basic content knowledge.

Thus, an obvious question to ask is: How well do textbook presentations fare compared to multimedia presentations in teaching introductory physics content?  The study reported here is an attempt to answer this question.  More specifically we compared three different modes of presenting the content of the first two weeks of a typical calculus-based introductory electricity and magnetism course: 1) Paper presentations based on a popular introductory textbook, 2) Web-based multimedia presentations designed according to principles from multimedia learning, and 3) Paper presentations consisting of the script from the web-based presentation in item 2. We were interested in comparing these three modes of presentation in terms of assessments of learning both immediately following the presentations, as well as after an extended period of time as a measure of retention.  In addition, we asked students two attitudinal questions regarding their use of textbooks in the course, and their preference for the materials used in this study compared to their current textbook.

\section{Description of the study}

\subsection{Choice of topics}

We specifically chose the material for our study from the first two weeks of an introductory, calculus-based electricity \& magnetism (E\&M) course. Topics included Coulomb's Law, superposition, electric fields from point charges and continuous distributions, electric flux, and Gauss' Law.  Students find E\&M concepts very abstract and difficult, making them ideal for investigating which type of presentation mode works best for helping students learn basic abstract content.

\subsection{Description of presentations}

\subsubsection{Web-based, multimedia learning modules}

Four pilot {\it multimedia learning modules} ({\it MLMs}), were developed: {\it 1)} Coulomb's Law, {\it 2)} Electric Fields, {\it 3)} Electric Flux, and {\it 4)} Gauss' Law.   The {\it MLMs} drew heavily on research related to multimedia learning summarized above.  Each {\it MLM} covered one lecture's worth of course content, and was divided into approximately 10 scenes. Each scene was implemented as a Flash\endnote{http://en.wikipedia.org/wiki/Adobe\_Flash} movie containing dynamic animations synchronized with an audio narration that was controlled by the student (pause, play, rewind, and position).  Embedded formative assessments were included in two or three of the scenes for each module, which consist of questions that must be answered correctly before the student can move to the next scene. The average narration time for an {\it MLM} was 12 minutes; students typically took about 17 minutes to complete a single {\it MLM} including the embedded assessment.\endnote{To experience these {\it MLMs} as would a student, please visit: http://research.physics.uiuc.edu/PER/demo\_iol\_212.html.}

\subsubsection{Scripts from multimedia learning modules}

One of the two paper-presentation modes consisted of scripts of the {\it MLM} narrations, (heretofore referred to as {\it MLM-script}) together with accompanying static pictures from the {\it MLMs}' animations.  Each {\it MLM-script} was about 7 pages long and contained 7 figures.\endnote{All of these scripts are accessible at: http://research.physics.uiuc.edu/PER/demo\_iol\_212.html.}

\subsubsection{Textbook-based materials}

The second paper-presentation mode consisted of verbatim material from a traditional textbook  (heretofore referred to as the {\it Textbook} group) subject to the constraint that the same worked out examples covered in the {\it MLMs} had to also be covered in the textbook-based materials.  The reason for this constraint is that the cognitive research literature suggests that students often focus on worked out examples to learn how concepts are instantiated\endnote{J. Sweller and G. A. Cooper, ``The use of worked examples as a substitute for problem solving in learning algebra,''  Cogn. \& Instr. {\bf 2}, 59--89 1985.; M. T. H. Chi, N. De Leeuw, M. H. Chiu, and C. Levancher, ``Eliciting self-explanations improves understanding,''  Cog. Sci. {\bf 18}, 439--477 1994.; M. T. H. Chi, M. Bassok, M. W. Lewis, P. Reimann, and R. Glaser, ``Self-explanations: How students study and use examples in learning to solve problems,''  Cog. Sci. {\bf 13}, 145--182 1989.}, and so we wanted to keep the worked out examples in all presentations equivalent. We selected materials from the Tipler and Mosca textbook\endnote{P. Tipler P. and G. Mosca, {\it Physics for Scientists and Engineers, Sixth Edition} (W. H. Freeman and Company, 2008).}, largely because the material and worked-out examples in this textbook paralleled those in the {\it MLMs}.  We deleted a total of six worked examples from the four textbook units that were not covered in the {\it MLMs}.  We also added one example from the Halliday, Resnick and Walker\endnote{D. Halliday, R. Resnick, and J. Walker, {\it Fundamentals of Physics} (John Wiley \& Sons, 2001).} textbook to match the one example in {\it MLM-script} that had no counterpart in the Tipler \& Mosca materials.  Each edited textbook unit was about 7 pages long and contained 8 figures and 2 examples.

\subsection{Study procedures}

Volunteers were solicited from students taking the first semester calculus-based mechanics course (Physics 211) at UIUC and offered compensation in exchange for participation in the study.  Students were paid \$15 for each of three 90-minute sessions, and a \$20 study-completion bonus. Students were randomly assigned to three groups, each receiving a different treatment ({\it MLM}, {\it MLM-script}, or {\it Textbook}).

The four units were delivered within two 90-minute sessions distributed over two days in the same week.  Each 90-minute session consisted of two ``lessons,'' each lasting about 45 minutes.  Each lesson began with a presentation (the {\it MLM} group sat at computers to receive the multimedia presentation; the other two groups received paper booklets containing either the {\it MLM-script}, or {\it Textbook} presentation) and was followed by a post-lesson assessment.  Two weeks later students returned for a retention test, which lasted about 1 hour.

\subsection{Assessments}

\subsubsection{Post-lesson assessments}

The Post-Lesson Assessments consisted of multiple-choice and free-response questions; to streamline this article and discussion of results, we will only present results from the multiple choice questions.  The multiple-choice questions were largely conceptual; some were taken from the Conceptual Survey or Electricity and Magnetism\endnote{D. P. Maloney, T. L. O'Kuma, C. J. Hieggelke, and A. Van Heuvelen, ``Surveying students' conceptual knowledge of electricity and magnetism'' Am. J. Phys {\bf 69}, S12--S23 2001.} and some were constructed by us.  The 43 multiple choice questions on the Post-Lesson Assessments may be obtained from the authors.

\subsubsection{Retention Test}

The Retention Test administered about two weeks following the lessons had 37 questions (both multiple choice and free-response), 33 of which were either identical or very similar (e.g., surface feature changes, change in order of answers) to questions on the Post-Lesson Assessments, and 32 of which were multiple choice.  Only results from the multiple choice questions will be presented here.  The 32 multiple choice questions from the Retention Test may be obtained from the authors.

\subsubsection{Survey questions}

At the end of the third ``retention'' testing session, the following two survey questions were administered to all three groups:

\begin{enumerate}
\item Compare the presentation of the material in this study, with that of your current textbook
   \begin{enumerate}
   \item Strongly prefer current textbook
   \item Somewhat prefer current textbook
   \item About the same
   \item Somewhat prefer these study materials
   \item Strongly prefer these study materials
   \end{enumerate}
\item How often do you now read your physics textbook?
   \begin{enumerate}
   \item Two or more times per week
   \item About once per week
   \item About 1 to 3 times per month
   \item Just to cram for exams
   \item Never
   \end{enumerate}
\end{enumerate}

\section{Results}

A total of 63 students who were enrolled in the calculus based mechanics course volunteered for the study and were randomly assigned to one of the three groups. The results reported here are based on the 45 students who completed the entire study (16 {\it MLM}, 13 {\it MLM-script}, 16 {\it Textbook}).  The average Physics 211 exam scores (based on combining scores on the three course hour exams) for each of the three groups, along with its uncertainty, are shown in the second column of Table~\ref{basicresults}. The spread is about what one would anticipate statistically, with a slight advantage to the {\it MLM-script} group. We will provide raw results, and results that use the covariance to adjust for differences between the groups as determined by these mechanics exam scores. 

First we will address how the three groups compare in terms of initial learning immediately following each lesson by analyzing the students' performance on the 43 multiple-choice questions of the four Post-Lesson Assessments. Then we will compare the three groups on the retention test administered two weeks following the lessons by analyzing the students' performance on the of 32 multiple choice questions from the Retention Test.

\subsection{Comparison of multimedia learning modules vs. textbook}

The combined results from the Post-Lesson Assessments and associated error are given in column 3 of Table~\ref{basicresults}. The {\it MLM} group had an average score of 75\% compared with the {\it Textbook} group's score of 63\%.  The standard deviation of the Post-Lesson Assessments scores is 13.5\%, which means students in the {\it MLM} group scored nearly a full standard deviation higher on the exams than their peers who used the textbook (effect size = 11.5/13.5 = 0.85).  Since the Textbook group's mechanics exam scores were slightly lower (1.5\%) than those of the {\it MLM} group we can use the correlation between the mechanics exam scores and the Post-Lesson Assessment scores to estimate an adjusted score for the Textbook group of 65\% as shown in column 4. Including this adjustment the {\it MLM} group still scored nearly 10 percentage points higher than the textbook group giving an effect size of about 0.7.  Performing an analysis of covariance based on the mechanics exam scores and the Post-Lesson Assessment scores, we obtain the result p $<$ 0.01\endnote{p represents the probability that the 10\% difference in exam scores between the two groups happened by random chance, assuming that the two groups are equivalent.}. Hence students using the {\it MLMs} demonstrated a large and statistically significant advantage in performance on the Post-Lesson Assessment items compared to students using the textbook. We now address the question of whether or not this advantage perseveres by looking at the Retention Test.

The Retention Test was given approximately two weeks after the lessons, and the raw results are shown in column 4 of Table~\ref{basicresults}.  Once again the {\it MLM} group performed about 13\% higher than the Textbook group, with an average score of 70\% compared with the Textbook group's raw average of 57\%, which gets adjusted to 58\% based on their mechanics hour exam performance. The adjusted effect size is 0.7 with p $<$ 0.01. Once again we have a large, statistically significant result showing that students using the {\it MLMs} not only learn more immediately following the lessons, but also retain that information better than the students using the textbook presentation two weeks following exposure.

\begin{figure}[h]
\caption{A Scatter Plot of a student's average mechanics exam score vs his/her combined Post-Learning and Retention Scores for both {\it MLM} and {\it Textbook} groups.\label{stelzerfig1}}
\includegraphics[width=15cm]{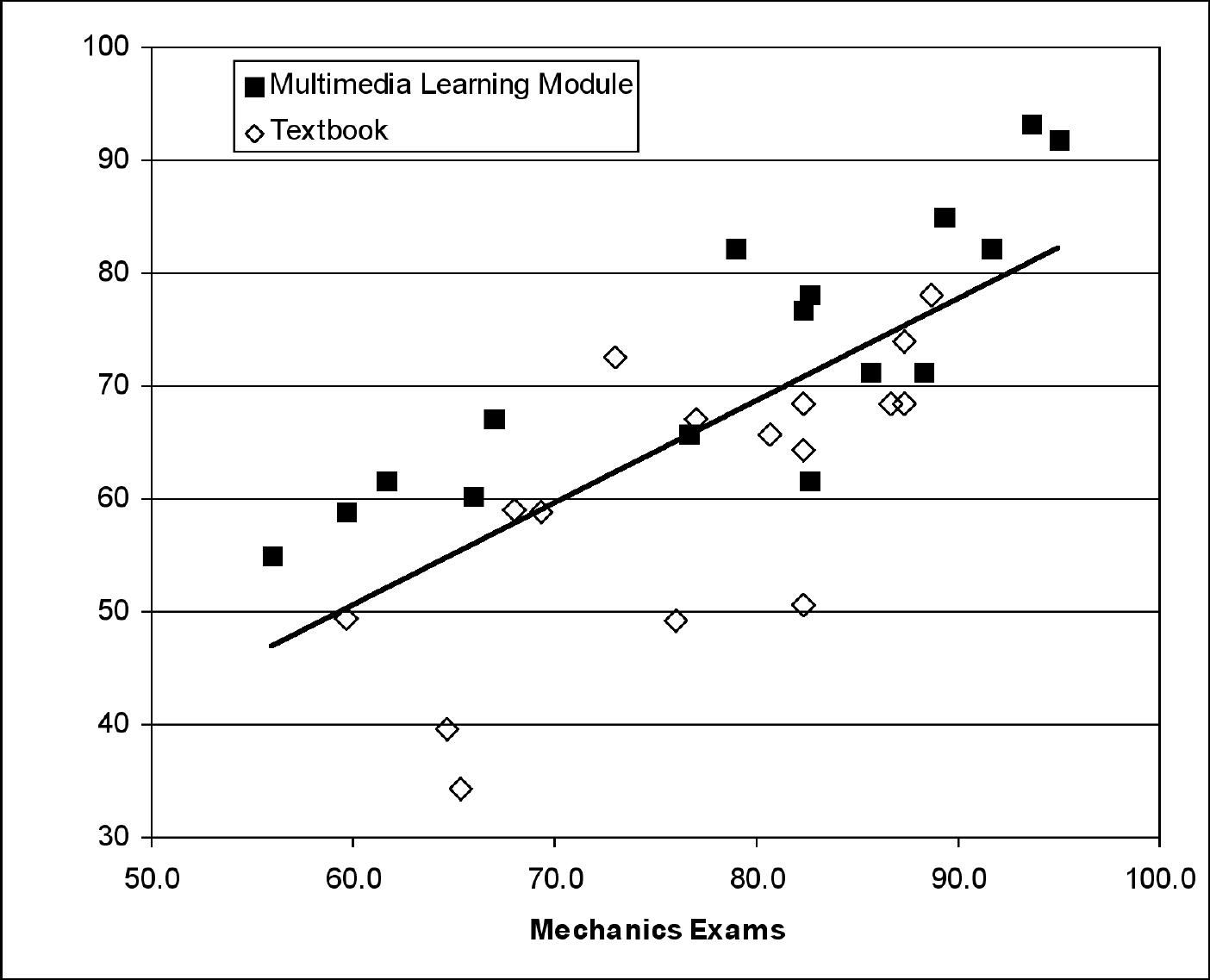}
\end{figure}

Figure~\ref{stelzerfig1} illustrates the significance of this effect. Each dot represents a student; on the horizontal axis is their average mechanics hour exam score, and on the vertical axis is their average combined score including both the Post-Lesson Assessments and the Retention Test (labeled Study Scores on the graph).  The line in the figure represents a linear fit to all of the data. Notice that 75\% of the {\it MLM} group scores are above this line, while 75\% of the Textbook group scores are below it. Hence, independent of the students' ability (based on their mechanics score), more learning and better retention resulted from the {\it MLM} presentations versus the {\it Textbook} presentation.

\subsection{Comparison of MLM-Scripts, with MLM and Textbook Presentations}

According to multimedia learning theory\endnote{See Ref.~\ref{mayer2001}.}, the large differential in performance in favor of the {\it MLMs} over the {\it Textbook} is likely attributable to two factors: 1) {\it MLMs} dispense information via {\it dual channels} (visual and auditory), and 2) {\it MLMs} create less memory load compared to the textbook. The motivation for including the {\it MLM-script} group in our study was to test the multimedia learning theory further by comparing its effectiveness to both the {\it MLM} and the {\it Textbook}.  The {\it MLM-script} presentation and the {\it Textbook} presentation differed in one important way according to multimedia learning theory: Multimedia learning theory states that presentation layout should minimize cognitive load by excluding extraneous words and pictures (the ``coherence principle''\endnote{See Ref.~\ref{mayer2001}.}); in short, a presentation should stay strictly ``on-message'' with little else to distract the learner from that message.  This principle, coupled with the ``dual channel'' principle, would predict that: 1) the {\it MLM-script} group should outperform the {\it Textbook} group (not only because the ``enhancements'' in textbooks serve to distract students from staying ``on-message,'' but also because the {\it MLM-scripts} were ``spartan'' presentations, with a tight story-line, augmented by the few crucial figures needed to illustrate the ideas in the text), but 2) The {\it MLM-script} group should perform lower than the {\it MLM} group (because, although both were ``on-message,'' the {\it MLM-scripts} lacked the dual channel processing available in the MLMs). 

\begin{figure}[h]
\caption{Average adjusted scores on the combined Post-Learning Assessments and the Retention Test for all three groups ({\it MLM}, {\it MLM-script}, and {\it Textbook}).\label{stelzerfig2}}
\includegraphics[width=15cm]{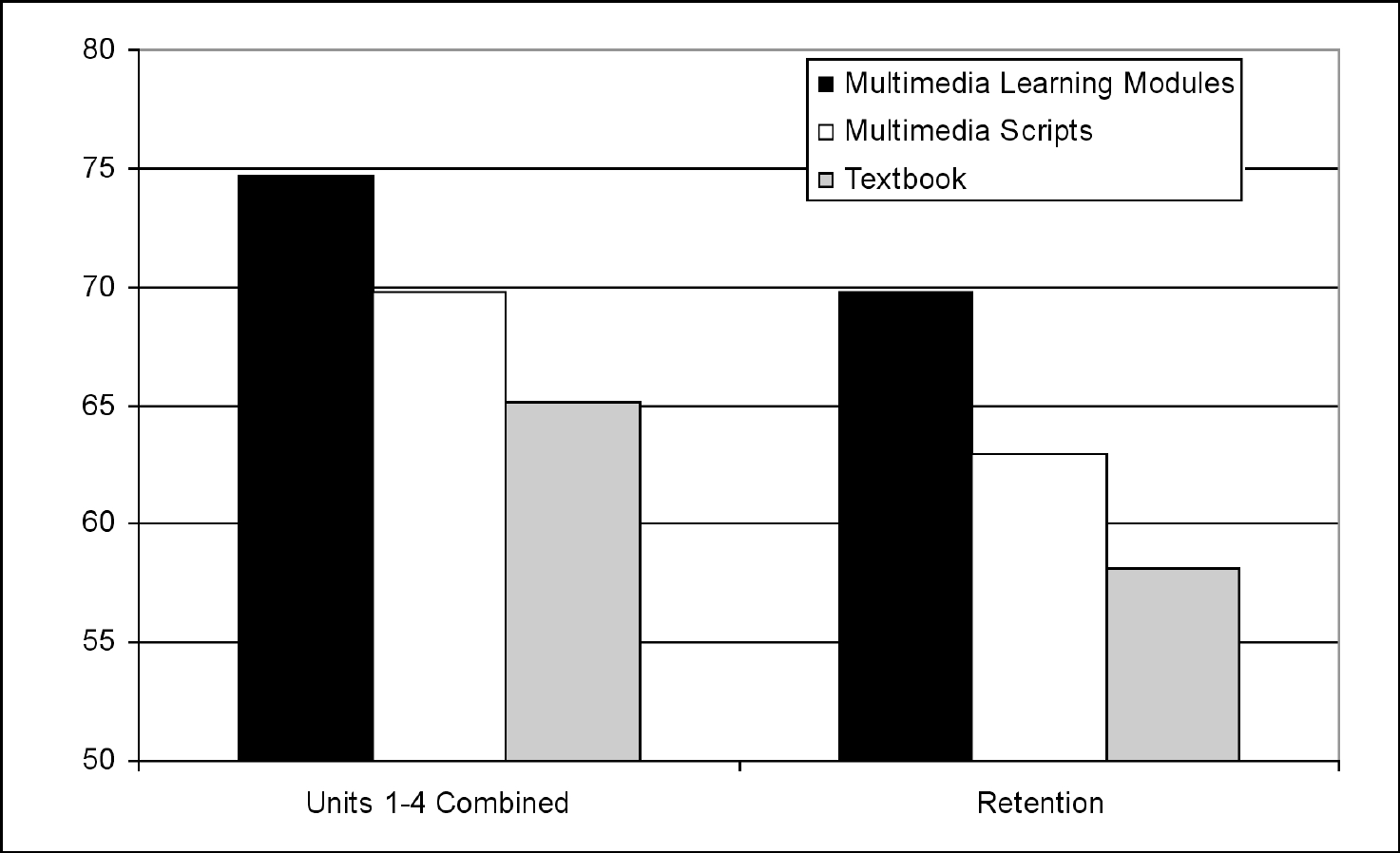}
\end{figure}

Our findings were in accord with these predictions, as shown in Table~\ref{basicresults} and Fig.~\ref{stelzerfig2}. The {\it MLM-script} group's scores are above those of the {\it Textbook} group, but below those of the {\it MLM} group.  Table~\ref{stats} shows the effect size $S$ and the p-value for all possible comparisons of the three groups for both the Post-Lesson Assessments and the Retention Test. The p values were computing using ANCOVA with the physics 211 exam scores as the covariant.

\subsection{Results of two survey questions\label{3C}}

\begin{figure}[h]
\caption{Survey results for frequency that student uses textbook in current physics course. Since exams occur about once per month, we have grouped choices c) 1-3 / month and  d) cram for exams into one category labeled every month.\label{stelzerfig3}}
\includegraphics[width=15cm]{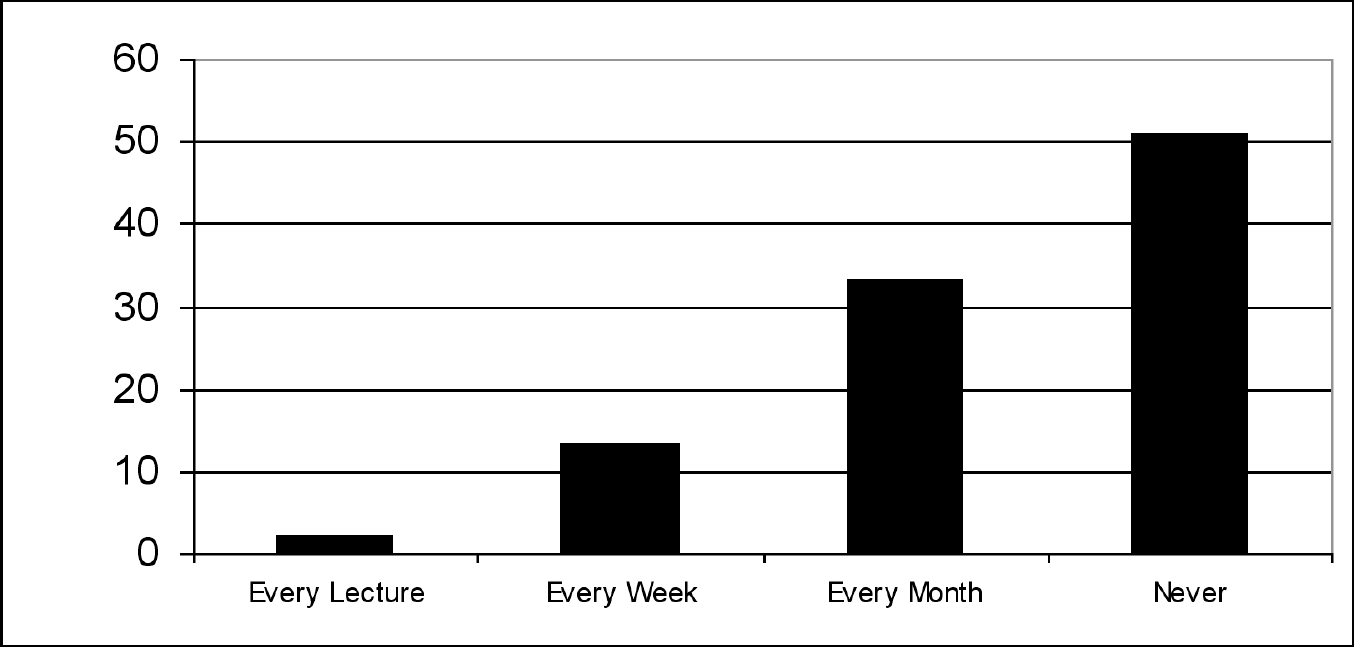}
\end{figure}

Table~\ref{surveyresults} and Fig.~\ref{stelzerfig3} summarize the results for the two survey questions.  We report the results of the first question for each group separately, and the second question for all students together (this second question did not depend on the treatment received; it only asked students to report on their use of the current textbook in their course).

As can be seen from the results of the second question summarized in Fig.~\ref{stelzerfig3}, students in this study reported similar usage of their textbooks as those in Table~\ref{textbooksurvey}.  Only 16\% of all students who finished the study reported that they read their textbooks once per week or more and 50\% claim to never have used the textbook the entire semester!  

The results from the first question support the statement made earlier, namely that today's generation of students prefer to learn from multimedia materials compared to textbooks.  The survey results for question 1 indicate that the {\it MLM} group showed a greater preference for the study materials encountered during the intervention over their traditional textbook as compared to the {\it Textbook} and the {\it MLM-script} groups (p $<$ 0.02, Fisher exact test).\endnote{The Fisher exact probability test is an alternative to the chi-squared test, suitable for situations where the basic assumptions of the chi-squared test are violated.  In our case, our small sample size renders the use of the chi-squared test invalid.}  (For this comparison the two text-based presentations were combined).  The response pattern between the {\it Textbook} and {\it MLM-script} groups did not significantly differ from each other (p = 0.4 Fisher exact test).

\section{General Discussion}

Our study has shown that learning of basic physics content by the current generation of undergraduates from a typical modern introductory textbook fared poorly when compared to learning from multimedia modules that were designed based on principles derived from research into multimedia learning.  The effect sizes obtained in this study are consistent with those from numerous other multi-media learning studies.  For example, Mayer and collaborators have performed a series of 39 experiments designed to compare the retention performance of groups using presentations that either were, or were not prepared according to the seven design principles they developed and they found positive effects in 33 of them\endnote{See Refs.~\ref{mayer2005} and \ref{mayer2001}.} with the average effect size being 0.8.  We conclude that multimedia learning modules represent a viable, and more easily enforceable alternative means of pre-lecture preparation for introductory physics students.

\begin{acknowledgments}

We wish to thank Adam Feil for his help in preparing this manuscript.

\end{acknowledgments}

\begingroup
 \theendnotes
\endgroup

\newpage

\section*{Tables}

\begin{table*}[!h]
\caption{Results of two questions from a survey administered in 2003, 2004 and 2005 to students completing introductory physics courses at UIUC\label{textbooksurvey}}
\begin{ruledtabular}
\begin{tabular}{p{3.5cm}p{3cm}p{4cm}p{3.5cm}}
& & {\it Algebra-based physics for life-sciences}& {\it Calc-based physics for engineers} \\
\hline
How often do you read the text before attending class? & Always \newline Regularly \newline Occasionally \newline Rarely \newline Never & 13 (2\%)\newline 70 (10\%)\newline 112 (17\%)\newline 151 (23\%)\newline 321 (48\%) & 32 (5\%)\newline 55 (9\%)\newline 116 (18\%)\newline 133 (21\%)\newline 297 (47\%) \\
\hline
Overall, how important was the text in helping you understand the course material? & Essential \newline Very Useful \newline Useful \newline Not Very Useful \newline Useless & 9 (1\%)\newline 31 (5\%)\newline 160 (24\%)\newline 254 (38\%)\newline 208 (31\%) & 30 (5\%)\newline 49 (8\%)\newline 114 (18\%)\newline 173 (28\%)\newline 256 (41\%) \\
\end{tabular}
\end{ruledtabular}
\end{table*}

\newpage

\begin{table*}[!h]
\caption{Performance of the three groups on hour exams, Post-Lesson Assessments, and the Retention Test.\label{basicresults}}
\begin{ruledtabular}
\begin{tabular}{p{3.5cm}ccccc}
 & \multicolumn{1}{p{3cm}}{Mechanics course exam scores} & \multicolumn{2}{p{4.5cm}}{Post-Lesson Assessments (\# of questions = 43)} & \multicolumn{2}{p{4.5cm}}{Retention Test (\# of questions = 32)} \\
 \hline
 & & Raw & Adjusted & Raw & Adjusted \\
\hline
{\it MLM} (n = 16) & $78.5 \pm 3.1$ & $74.7 \pm 2.6$ & 74.7 & $69.7 \pm 4.0$ & 69.8 \\
{\it Textbook} (n = 16) & $76.9 \pm 2.3$ & $63.2 \pm 3.1$ & 65.2 & $56.8 \pm 4.1$ & 58.1 \\
{\it MLM-script} (n = 13) & $81.4 \pm 3.3$ & $71.7 \pm 3.0$ & 69.7 & $65.1 \pm 3.7$ & 63.0 \\
\end{tabular}
\end{ruledtabular}
\end{table*}

\newpage

\begin{table*}[!h]
\caption{Effect size (S) and significance (p) for comparisons of the performance of the three groups on both Post-Lesson Assessments and the Retention Test.\label{stats}}
\begin{ruledtabular}
\begin{tabular}{p{2.5cm}cccc}
 & \multicolumn{2}{c}{Post-Lesson Assessments} & \multicolumn{2}{c}{Retention Test} \\
 \hline
 & {\it MLM} & {\it MLM-script} & {\it MLM} & {\it MLM-script} \\
 \hline
{\it Textbook} & S = 0.74, p $<$ 0.01 & S = 0.36 p = 0.06 & S = 0.67, p $<$ 0.01 & S = 0.31, p = 0.16 \\
{\it MLM-script} & S = 0.46, p = 0.03 & & S = 0.45, p = 0.04 & \\
\end{tabular}
\end{ruledtabular}
\end{table*}

\newpage

\begin{table*}[!h]
\caption{Results of survey questions.\label{surveyresults}}
\begin{ruledtabular}
\begin{tabular}{cccc}
\multicolumn{4}{p{15cm}}{Survey Question 1: Compare the presentation of the material in this study, with that of your current textbook\newline
a) Strongly prefer current textbook\newline
b) Somewhat prefer current textbook\newline
c) About the same\newline
d) Somewhat prefer these study materials\newline
e) Strongly prefer these study materials} \\
\hline
Response & {\it MLM} & {\it MLM-script} & {\it Textbook} \\
\hline
a), or b), or c) & 3 & 8 & 8 \\
d), or e) & 13 & 5 & 8 \\
\end{tabular}
\end{ruledtabular}
\end{table*}

\newpage

\section*{Figure Captions}

Figure 1:  A Scatter Plot of a student's average mechanics exam score vs his/her combined Post-Learning and Retention Scores for both {\it MLM} and {\it Textbook} groups.

Figure 2:  Average adjusted scores on the combined Post-Learning Assessments and the Retention Test for all three groups ({\it MLM}, {\it MLM-script}, and {\it Textbook})

Figure 3:  Survey results for frequency that student uses textbook in current physics course. Since exams occur about once per month, we have grouped choices c) 1-3 / month and  d) cram for exams into one category labeled every month.

\newpage

\section*{Figures}

\newpage

\newpage

\end{document}